\documentclass[]{aipproc}
\layoutstyle{6x9}
\usepackage{epsfig}
\newcommand{\be}{\begin{equation}}
\newcommand{\ee}{\end{equation}}
\newcommand{\bea}{\begin{eqnarray}}
\newcommand{\eea}{\end{eqnarray}}
\newcommand{\w}{\omega}
\newcommand{\G}{\Gamma}

\newcommand{\de}{\delta}
\newcommand{\De}{\Delta}
\newcommand{\ga}{\gamma}
\newcommand{\ro}{\rho}
\newcommand{\la}{\lambda}

\newcommand{\ov}[1]{\overline{#1}}
\def\slr#1{\setbox0=\hbox{$#1$}           
   \dimen0=\wd0                                 
   \setbox1=\hbox{/} \dimen1=\wd1               
   \ifdim\dimen0>\dimen1                        
      \rlap{\hbox to \dimen0{\hfil/\hfil}}      
      #1                                        
   \else                                        
      \rlap{\hbox to \dimen1{\hfil$#1$\hfil}}   
      /                                         
   \fi}
\begin{document}
\title{Recent Developments in the Bethe-Salpeter Description of Light Mesons}
\author{Peter Watson}{
 address={Institute for Theoretical Physics, University of Giessen, Heinrich-Buff-Ring 16, D-35392 Giessen, Germany},
}
\begin{abstract}
Results for the light meson mass spectrum from a Bethe-Salpeter approach are presented.  The results obtained in the standard framework are Poincar\'{e} covariant and compare favourably with lattice results.  Using a more sophisticated scheme, the pseudoscalar, vector and $1^{++}$ ($a_1/f_1$) axialvector charge eigenstate masses are unaltered whereas the $1^{+-}$ ($b_1/h_1$) axialvector meson mass is raised.
\end{abstract}
\maketitle
The Bethe-Salpeter equation [BSE] is the fully relativistic description of the two-body bound state problem.  It has been found in the last decade that the ladder truncation of the BSE, using as input the quark propagators derived from the rainbow truncation of the Schwinger-Dyson equation [DSE], gives rise to a good description of the light flavor non-singlet pseudoscalar and vector mesons \cite{Jain:1993qh}.  The underlying mechanism for this is chiral symmetry manifested through the flavor non-singlet axialvector Ward-Takahashi identity [AXWTI].  By ensuring that the kernels of both equations respect the AXWTI, it is shown that the pion emerges as both a bound state of massive constituents and as an almost massless Goldstone boson of the broken chiral symmetry \cite{Maris:1997hd}.

The ladder truncation of the homogeneous BSE for quark-antiquark mesons is written (working in Euclidean space with Hermitian Dirac matrices obeying $\{\ga_{\mu},\ga_{\nu}\}=2\de_{\mu\nu}$):
\be
\G(p;P)=-\frac{4}{3}\int\frac{d^4k}{(2\pi)^4}g^2\De_{\mu\nu}(p-k)\ga_{\mu}S(k_+)\G(k;P)S(k_-)\ga_{\nu}~~~,
\label{eq:bse1}
\ee
where $\G$ is the Bethe-Salpeter amplitude, $k_+=k+\xi P$ and $k_-=k+(\xi-1)P$ with $\xi=[0,1]$ the momentum sharing parameter between the two quarks.  Invariance of the resulting observables with respect to $\xi$ is a reflection of Poincar\'{e} covariance.  The total momentum $P=p_+-p_-$ is such that the equation is solved for $P^2=-M^2$ where $M$ is the mass of the meson.  In Eq.~(\ref{eq:bse1}), the dressed quark propagators are the solution of the rainbow quark DSE
\be
S^{-1}(p)=\imath\slr{p}+m+\frac{4}{3}\int\frac{d^4k}{(2\pi)^4}g^2\De_{\mu\nu}(p-k)\ga_{\mu}S(k)\ga_{\nu}~~~,
\ee
where $m$ is the current mass parameter of the quark; the two truncations being consistent with the AXWTI.  The effective interaction $g^2\De_{\mu\nu}(q)$ has the following form \cite{Alkofer:2002bp}:
\be
g^2\De_{\mu\nu}(q)=t_{\mu\nu}(q)4\pi^2D\frac{q^2}{\w^2}\exp{\left(-\frac{q^2}{\w^2}\right)}
\ee
($t_{\mu\nu}$ is the transverse projector).  The two parameters $\w$ and $D$ set the length scale and magnitude of the interaction.  The integrals are UV convergent and all renormalisation constants are unity.  The BSE is solved by writing the amplitude $\G$ in its most general form consistent with the desired parity and charge conjugation properties and expanding the resulting scalar functions as a Chebyshev series in the angular variable.  Numerical results are shown in Table~\ref{tab:numres1} using the parameters $\w=0.5GeV$, $D=16GeV^{-2}$, $m_u=m_d=5MeV$, $m_s=115MeV$.  One can see that the pseudoscalar and vector meson results are in good agreement, but the axialvector mesons are $\sim300MeV$ too light.  Varying the momentum sharing parameter ($\xi$) one finds that for sufficient numbers of Chebyshev moments ($N\geq6$), the mass results are stable, demonstrating the Poincar\'{e} covariance \cite{Alkofer:2002bp}.

\begin{table}[h]
\caption{\label{tab:numres1}Mass results (in $GeV$) for the light meson masses using the rainbow/ladder truncation.}
\begin{tabular}{ccccccc}\hline
$J^{PC}$&\multicolumn{3}{c}{$M_{BS}$}&\multicolumn{3}{c}{$M_{exp}$}\\ \hline
&$\ov{u}u$&$\ov{u}s$&$\ov{s}s$&$\ov{u}u$&$\ov{u}s$&$\ov{s}s$\\\hline
$0^{-+}$ & 0.137 & 0.492 & --    & 0.135 & 0.498 & -- \\
$1^{--}$ & 0.758 & 0.946 & 1.078 & 0.770 & 0.892 & 1.020 \\
$1^{+-}$ & 0.915 & 1.075 & 1.233 & 1.230 & 1.270 & 1.170? \\
$1^{++}$ & 0.936 & 1.075 & 1.291 & 1.230 & 1.270 & 1.282  \\ \hline
\end{tabular}
\end{table}

\begin{figure}[t]
{\epsfig{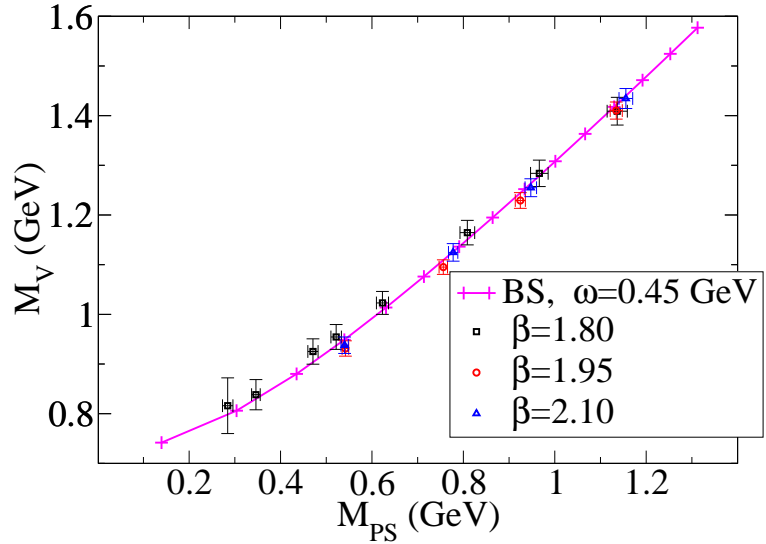}}
\caption{\label{fig:latcomp}Comparison of BSE and lattice results for pseudoscalar and vector meson masses.  The lattice data are from CP-PACS (unquenched) \cite{AliKhan:2001tx}.}
\end{figure}

It is interesting to compare the BSE results for the pseudoscalar and vector masses with lattice results at different quark masses.  Figure~\ref{fig:latcomp} shows a comparison of $M_V$ vs. $M_{PS}$ using unquenched CP-PACS data \cite{AliKhan:2001tx} and the BSE results with $\w=0.45GeV$, $D=24.4GeV^{-2}$.  There is good agreement between the lattice and the BSE results over a wide range of $M_{PS}$.

To improve the description of the axialvector mesons one must augment the truncation scheme to include nontrivial vertex corrections whilst maintaining the AXWTI.  As an initial attempt we consider an abelian, one-loop correction to the quark-gluon vertex with the following form \cite{Watson:2004kd}:
\be
\G_{\mu}(k,p)=\ga_{\mu}+\frac{1}{6}\int\frac{d^4q}{(2\pi)^4}\ga_{\ro}S(k+q)\ga_{\mu}S(p+q)\ga_{\la}g^2\ov{\De}_{\ro\la}(q)~~~.
\ee
To make the system tractable, the interaction in the vertex correction is taken as $g^2\ov{\De}_{\ro\la}(q)=(2\pi)^4G\de^4(q)t_{\ro\la}(q)$ which reduces the integral to an algebraic expression.  The parameter $G$ is tuned such that the integrated strengths of $\De$ and $\ov{\De}$ are equal.  At spacelike momenta the quark propagator is only slightly modified from the rainbow case but this is not true at general complex momenta due to the different analytic structures introduced by the $\de$-function.  As before, one can construct an appropriate kernel for the BSE which preserves the AXWTI and the charge conjugation symmetry.  The results show that the charge eigenstate pseudoscalar, vector and $1^{++}$ ($a_1/f_1$) axialvector meson masses are largely unaltered but the $1^{+-}$ ($b_1/h_1$) axialvector mass increases $\sim300MeV$, comparing reasonably with the experimental observation \cite{Watson:2004kd}.

To summarize, the BSE -- when using input from the quark DSE whilst respecting the AXWTI -- provides a powerful framework to study the meson mass spectrum.  There are two main areas with which to proceed: the first being the inclusion of more sophisticated truncations of the kernels; an exploratory attempt being described here.  The second area to investigate is to include a mechanism for dynamical meson decay and multiquark states.  Such an effort has been started in ref.~\cite{Watson:2004jq}.

\vspace{0.5cm}
\noindent\footnotesize{It is a pleasure to thank N. Brambilla and the organizing committee for an enjoyable and productive conference.  This work was supported by FZ J\"{u}lich.}


\end{document}